\documentclass[%
 reprint,
 amsmath,amssymb,
 aps,
]{revtex4-2}

\usepackage{graphicx}
\usepackage{dcolumn}
\usepackage{bm}

\begin{document}

\preprint{APS/123-QED}

\title{Few-photon single ionization of cold rubidium in the over-the-barrier regime}
\thanks{wangxch1@shanghaitech.edu.cn;\\ ye\_difa@iapcm.ac.cn;\\ jiangyh@sari.ac.cn}%

\author{Huanyu Ma,$^{1,2,3}$ Xincheng Wang,$^{1,*}$ Linxuan Zhang,$^{4,5}$ Zhihan Zou,$^{1,2,3}$ Junyang Yuan,$^{1,2,3}$ Yixuan Ma,$^{1,2,3}$ Rujin Lv,$^{2,3}$ Zhenjie Shen,$^2$ Tianmin Yan,$^2$ Matthias Weidem\"uller,$^{6}$ Difa Ye,$^{4,*}$ and Yuhai Jiang $^{1,2,3,*}$}
\affiliation{%
$^1$Center for Transformative Science and School of Physical Science and Technology, ShanghaiTech University, Shanghai 201210, China;\\
$^2$Shanghai Advanced Research Institute, Chinese Academy of Sciences, Shanghai 201210, China\\
$^3$University of Chinese Academy of Sciences, Beijing 100049, China\\
$^4$Laboratory of Computational Physics, Institute of Applied Physics and Computational Mathematics, Beijing 100088, China\\
$^5$Graduate School, China Academy of Engineering Physics, Beijing 100193, China\\
$^6$Physikalisches Institut, Universit\"at Heidelberg, Im Neuenheimer Feld 226, 69120 Heidelberg, Germany
}%

\date{\today}

\begin{abstract}
Photoionization of the rubidium (Rb) atoms cooled in a magneto-optical trap, characterized by the coexistence of the ground 5$S_{1/2}$ and the excited 5$P_{3/2}$ states, is investigated experimentally and theoretically with the 400 nm femtosecond laser pulses at intensities of $I=3\times10^9$ W/cm$^2$ $-4.5\times10^{12}$ W/cm$^2$. Recoil-ion momentum distribution (RIMD) of Rb$^+$ exhibits rich ring-like structures and their energies correspond to one-photon ionization of the 5$P_{3/2}$ state, two-photon and three-photon ionizations of the 5$S_{1/2}$ state, respectively. With the increasing of $I$, we find that experimental signals near zero-momentum (NZM) in RIMDs resulted from the 5$P_{3/2}$ state enhance dramatically and its peaked Rb$^+$ momenta dwindle obviously while that from the 5$S_{1/2}$ state is maintained. Meanwhile, the ion-yield ratio of the 5$S_{1/2}$ over the 5$P_{3/2}$ states varies from $I$ to $I^{1.5}$ as $I$ increases. These features indicate a transition from perturbative ionization  to strong-perturbative ionization for the 5$P_{3/2}$ state. Numerical simulations by solving the time-dependent Schr\"odinger equation (TDSE) can qualitatively explain the measurements of RIMD, photoion angular distributions, as well as ion-yield ratio. However, some discrepancies still exist, especially for the NZM dip, which could stem from the electron-electron correlation that is neglected in the present TDSE simulations since we have adopted the single-active-electron approximation.


\end{abstract}


\maketitle


\section{\label{sec:level1} INTRODUCTION }

The study of the ionization of materials in various light sources can be traced back to the photoelectric effect, which is the most elementary process in light-matter interactions~\cite{Photoelectriceffect,Photoelectriceffect2}. With the development of ultra-short super-intense laser, a series of novel nonlinear physical phenomena such as the multi-photon ionization, tunneling ionization, and over-the-barrier ionization are revealed~\cite{reviewliterature}. These different ionization mechanisms are distinguished well via the Keldysh parameter $\gamma =\sqrt{I_{p}/2U_{p}}$~\cite{Keldysh1964}, in which $I_p$ is the ionization energy of an atom and $U_{p}={2\pi I}/{c \omega^2}$ is the average quiver energy of an electron in a laser field with intensity $I$ and frequency $\omega$ , $c$ is the vacuum light speed in atomic units. When the laser intensity is low or the laser frequency is high, for $\gamma\gg1$, photoionization is considered to be a perturbative process described by the absorption of multiple photons~\cite{MPI}. In the perturbative multiphoton ionization (MPI) regime, the ionization rate depends on the laser intensities according to a power law $\mathcal{Y}\sim \emph{I}^n$ with $n$ of the number of photons absorbed~\cite{yield}. Photoelectron spectra can show a series of equal-spaced sharp peaks separated by one photon energy and these corresponding peaks are so-called above threshold ionization (ATI)~\cite{ATI1,ATI2}. The spectral strength of ATI decreases rapidly with the number of absorbed photons. As $\emph{I}$ increases, the strength of ATI peaks does not well follow $\mathcal{Y}\sim \emph{I}^n$ and the position of ATI is shifted to the lower values by $E_k=n\hbar \omega -I_p-U_p$, where the effect of $U_p$ becomes visible, resulted in strong-perturbative MPI phenomenon~\cite{Up,Up2}. As $\emph{I}$ increases continually, the ionization mechanism of the electrons will fall into the tunneling ionization~\cite{TI,TI2,TI3}, in which the laser field distorts the atomic potential to form a potential barrier through which the electron can tunnel. In this case,  $\gamma\ll1$ and the energy spectrum resulted from tunneling ionization is much smoother than that of MPI.
At $I_{\rm{OBI}}=c\mathcal{E}_{\rm{OBI}}^2/8\pi$ with $\mathcal{E}_{\rm{OBI}}=I_p^2/4Z$~\cite{OBI,OBI2}, the barrier formed by the Coulomb potential and the laser field is completely suppressed, and the electrons can classically escape via the over-the-barrier ionization (OBI). In general, the aforementioned ionization pictures categorized by the values of $\gamma$ and $\mathcal{E}_{\rm{OBI}}$ work quite well for noble gases, where MPI and OBI are separated by tunneling ionization at two different regimes of laser intensity~\cite{NOBIT}.

For alkali atoms, $I_{\rm{OBI}}$ is relatively small due to its lower ionization threshold in comparison with noble gases and its corresponding Keldysh parameter is far larger than $1$, which leads to the overlapping of the MPI and the OBI regimes. Some analysis have shown that when the laser intensity exceeds the OBI threshold, due to strong depletion of the ground state, most atoms can be ionized before reaching the peak intensity of the laser. It makes not meaningful to use the Keldysh parameter to classify tunneling and multiphoton ionization for the lithium atoms~\cite{OBIshort,Lir}. In the literature, sustained MPI beyond the OBI was previously measured lithium atoms and sodium atoms are both in the ground state, and theoretically studied in potassium~\cite{Nalilun,Li,Na,K}. Later, Wessels \emph{et al}.~\cite{Rb} measured strong-field ionization probabilities of ultra-cold Rb atoms and concluded that MPI remains a dominating mechanism even when $I>I_{\rm{OBI}}$. So far, most of experimental and theoretical studies for alkali atoms focused on the ionization mechanisms of the ground state with various photon wavelengths.


For cold Rb atoms cooled in a magneto-optical trap (MOT), concerned in this article, there is natural coexistence of the ground 5$S_{1/2}$ and the excited 5$P_{3/2}$ states.  In present mixture regime of MPI and OBI, ionization mechanisms will behave differently along with change of laser intensities because of the difference of ionization energies of 5$S_{1/2}$ and 5$P_{3/2}$ states, which might manifest in the variation of the ionization probabilities and the associated RIMDs as a function of laser intensity.   Besides, photoion augular distributions (PAD), another particularly sensitive observable, can provide an additional view of the underlying mechanisms involved in photoionization and their related properties, such as information on continuous states and inter-channel coupling~\cite{LiPAD,LiPAD2,NaPAD}. 

Here, we apply a 400 nm linearly polarized femtosecond laser to study the single photonionization of cold rubidium atoms created with MOT. We investigate the ion-yield and the RIMDs of Rb$^+$  at laser intensities of $I=3\times10^9-4.5\times10^{12}$ W/cm$^2$ penetrating across the OBI region from the MPI regime. The magneto-optical trap recoil ion momentum spectroscopy (MOTRIMS)~\cite{MOTRIMS1,MOTRIMS2,MOTRIMS3,MOTRIMS4,junyang,yuan2020}, combining cold atoms, strong laser pulse, and ultrafast technologies, was employed to detect Rb$^+$, in which the rubidium targets are cooled down to hundreds of $\mu K$ in order to achieve high resolution recoil-ion spectroscopy~\cite{Lirenyuan}. 
Owing to the laser wavelength that we have chosen, different ionization pathways resulted from the Rb(5$S_{1/2}$) and Rb(5$P_{3/2}$) states are well separated in the RIMDs as concentric rings. It delivers possibilities for comparative studies of the different ionization mechanisms of Rb(5$S_{1/2}$) and Rb(5$P_{3/2}$) under the same laser parameters. As we will show, the relatively strength of these few- (one-, two-, or three-) photon ionization (FPI) rings can be used to determine the population ratio of the excited state over the ground state and as an indication of the onset of strong-perturbative FPI. Moreover, the PAD of each ring can be also extracted, which provides quantitative information on the dominant partial waves linking to the so called ``asymmetry parameters" of the PAD. Finally, we point out that although most of the experimental measurements can be well explained by the \emph{ab-initio} simulations based on solving the time-dependent Schr\"odinger equation (TDSE), there still exist some differences in detail, especially for the intensity dependence of the near-zero momentum dip in the momentum spectra. Possible reasons for the discrepancy have been discussed in detail. Our benchmark measurements can serve as a stringent test ground for the further development of theory.

The paper is organized as follows: The experimental devices and theoretical simulation methods are introduced in Secs. II and III, respectively. Main results are demonstrated and discussed in detail in Sec. IV. Finally, we draw a conclusion in Sec. V. Atomic units (a.u.) are used throughout the paper, unless otherwise specified.

\section{EXPERIMENT}
\begin{figure}[tbp]
\centering
\includegraphics[width=8.5cm,height=7.5cm]{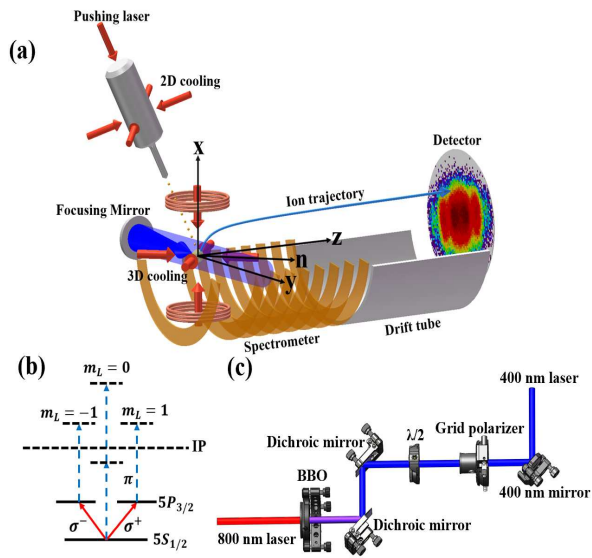}
\caption{\label{fig1}(a) Schematic diagram of the MOTRIMS device. The Rb atoms are pre-cooled in a 2D MOT, then pushed by a red-detuned pushing laser into the main experimental chamber, where they are cooled again by the 3D MOT and trapped in the center of the cavity, at last ionized by a 400 nm femtosecond laser. The red arrows represent the 780 nm cooling lasers, and the blue arrow represents the 400 nm femtosecond laser focused by the spherical on-axis concave mirror. We define the polarization direction of the femtosecond laser as $z$ and its propagation direction as $y$. There are six cooling laser beams: two in the $x$ axis and four in the $yz$ plane spanning an angle of $45^{\circ}$ with respect to the $z$ axis. (b) Diagram of the energy levels of the Rb atom. The red arrows represent the 780 nm cooling laser which prepares the Rb atom in the exited state, and the blue dashed arrows represent the 400 nm ionizing femtosecond laser. (c) Layout of the beamline for the generation of a 400 nm femtosecond laser pulse. The colors red, purple, and blue represent an 800 nm seed laser beam, the 400 nm and 800 nm co-beam, and the 400 nm final output laser beam, respectively. }
\end{figure}

A schematic diagram of the experimental devices, the MOTRIMS, is shown in Fig.~\ref{fig1}(a). Since the details of this setup can be found in Ref~\cite{Lirenyuan}, only a brief description is presented here. An intense femtosecond laser is focused on the target by a spherical mirror with a focal length of about 75 mm. The recoil ions are accelerated by a uniform electric field ($\sim$ 0.5 V/cm) and then guided through a field-free drift region. The length of acceleration region and drift region are 12 cm and 68 cm respectively. The arriving time and impact position of the ions are measured by a time-and-position sensitive detector, a microchannel plate (MCP) chevron stack with a delay-line anode, to reconstruct the initial momentum vectors.

Cold atoms used in this work originate from Rb vapor, which is pre-cooled by a two-dimensional (2D) MOT in a glass cavity with ultra-high vacuum. A red-detuned light is used to push these pre-cooled atoms to the main experimental chamber, where they can be recaptured and further cooled by a three-dimensional (3D) MOT. It should be noted that the cold Rb targets can be prepared in three different ways in the reaction regime. With 3D MOT cooling laser and gradient magnetic field on it is depicted as 3D MOT target, while by switching off the magnetic field, the Molasses target can be prepared. Furthermore, the direct Rb beam from 2D MOT, which has the lowest density, can also be used without further cooling, this is referred as 2D MOT target. With the 780 nm circularly-polarized cooling lasers used in the 2D and 3D MOT, the Rb atoms can be excited from the ground state (5$S_{1/2}$) to 5$P_{3/2}$ state. The energy levels of Rb atom are shown in Fig.~\ref{fig1}(b). The population of the ground state is estimated to be several times higher than the excited state for all three targets, depending on the intensity and detuning of the cooling laser~\cite{ratio1,ratio2,ratio3,ratio4}. It should be noted that, since the temperatures of these three targets are all on the order of 100 $\mu$K, so the influence of thermal motion on the recoil ion momentum can be neglected.

The laser pulse used in the experiment is produced by a Ti:sapphire femtosecond laser operating at 1 KHz, with a central wavelength of 800 nm and a pulse duration (full width at half maximum, FWHM) of 35 femtoseconds. As shown in Fig.~\ref{fig1}(c), a 200-$\mu$m thick $\beta$-BBO (beta-barium borate) octave crystal generates the second harmonic (400 nm) of the fundamental frequency laser (800 nm) with conversion efficiency as high as $\sim30\%$. With two dichroic mirrors (98$\%$ reflectance at 400 nm and 99$\%$ transmittance at 800 nm), pure 400 nm laser can be obtained, which is used to ionize the Rb target in the reaction chamber. The intensity of the beam is controlled by the $\lambda$/2 wave plate in front of a grid polarizer. In this work, the laser peak intensities are estimated to be $I=3\times10^9-4.5\times10^{12}$ W/cm$^2$. It should be noted that the grid polarizer ensure a stable linear polarization along horizontal direction.

\section{THEORETICAL METHOD}
Theoretically, the system is described by the TDSE in the length gauge and under the dipole approximation
\begin{equation}
i \frac{\partial \Psi ({\bf{r}}, t)}{\partial t}  =  \left[ -\frac{1}{2} \nabla^2 + V_0 ({r}) + V ({\bf{r}}, t) \right]\Psi ({\bf{r}}, t).
\label{eq:one}
\end{equation}
Here $V_{0}(r)$ is the effective atomic potential of Rb~\cite{Rbpotential} and $V(\bf{r},t)$ stands for the atom-field interaction:
\begin{gather}
V_0 ({r})  =  - \frac{1 + (Z - 1) e^{-a_1 r} + a_2 re^{-a_3
  r}}{r},\\
V ({\bf{r}}, t)  =  \mathcal{E} (t) r \cos \theta,
\end{gather}
where $\bf{r}$ is the position of electron with respect to the nucleus and $\theta$ the angle between $\bf{r}$ and the polarization direction ($z$ axis) of the laser electric field, and the atomic parameters are $Z=37$, $a_{1}=3.431$, $a_{2}=10.098$, and $a_{3}=1.611$. The external electric field $\mathcal{E}(t)$ takes the form
\begin{equation}
\mathcal{E} (t) = \mathcal{E}_0 \sin^2 \left( \frac{\pi t}{2 \tau}
  \right) \cos (\omega t),
\end{equation}
in which $\mathcal{E}_{0}$ represents the electric field amplitude, and $\tau$ is the pulse duration defined as the full-width at half maximum (FWHM). The TDSE is solved by means of the generalized pseudospectral technique in the spherical coordinates~\cite{tong,wuxiaoxia}. Then, to get the momentum resolved ionization probability, we expand the final wave function in the momentum-normalized Coulomb wave functions~\cite{Coulombwave}
\begin{equation}
\Psi^C_{{\bf{p}}} ({\bf{r}})  = \sum_{l m}
  \sqrt{\frac{2}{\pi} } i^l e^{- i (\delta_l + \vartriangle_l)}  \frac{R_{pl} ({r})}{pr} Y_{l m}^{\ast} (\hat{\bf{p}}) Y_{l m}
  (\hat{\bf{r}}),
\end{equation}
with $l$ the orbital quantum number, $m$ the magnetic quantum number, $\delta_{l}$ the phase shift caused by the short-range distortion of the asymptotic Coulomb field, $\triangle_{l}$ the Coulomb phase shift, $Y_{lm}$ the spherical harmonics, and $R_{pl}$ the reduced radial function satisfies the equation:
\begin{equation}
- \frac{1}{2}  \frac{d^2 R_{pl}}{d r^2} + \left[ \frac{l (l + 1)}{2 r^2} +
  V_0 (r) \right] R_{p l}  =  \frac{1}{2} p^2 R_{p l}.
\end{equation}
The momentum distribution of the photoelectron is then determined by
\begin{equation}
f ({\bf{p}}) = \left| \sum_{l m} \frac{1}{p}  \sqrt{\frac{2}{\pi} }
  i^{- l} e^{i (\delta_l + \vartriangle_l)} Y_{l m} (\hat{\bf{p}})
  \int^{\infty}_0 R_{p l} (r) \chi_{l m} (r) d r \right|^2,
\end{equation}
in which $\chi_{lm}(r)$ is the reduced radial function of the final wave function corresponding to the spherical harmonic $Y_{lm}$.

In our simulations, the time evolutions starting from the ground state and the excited state need to be traced separately and then the final momentum distributions should be summed together incoherently. For convenience and without loss of generality, we assume that the femtosecond laser propagates along the $y$ axis, with its electric field polarizes in the $z$ direction, as shown in Fig.~\ref{fig1}(a). Meanwhile, there exit six beams of cooling laser which will excite and polarize the electrons in different directions, i.e., along the cooling laser beam directions $n_1$ ,$n_2$, $n_3$, $n_4$, $n_{5}(+x)$ and $n_{6}(-x)$, with a magnetic quantum number $m=1$. Therefore, the initial wavefunction of the excited electrons can be expressed as:
\begin{equation}
 \psi_{5 P}^{n_i} (r, \theta, \varphi)  =  R_{5 P}  (r) Y^{n_i}_{11} (\theta,
  \varphi).
\end{equation}
It should be noted that the polarization directions $n_i$ $(i=1, 2, \cdots, 6)$ of the excited electrons are different from the main axis of the system that is afore chosen along the ionizing laser's polarization direction $z$. Thus it would be convenient to transform the spherical harmonic functions as follows:
\begin{eqnarray}
 &Y&^x_{11}  =   \frac{1}{2} Y_{11}^z + \frac{\sqrt{2}}{2} Y_{10}^z +
  \frac{1}{2} Y_{1 - 1}^z, \\
 & Y&^{- x}_{11}  =  \frac{1}{2} Y_{11}^z - \frac{\sqrt{2}}{2} Y_{10}^z
  + \frac{1}{2} Y_{1 - 1}^z, \\
 & Y&^{n_1}_{11}  =   \frac{2+\sqrt{2}}{4} Y_{11}^z + \frac{i}{2}
  Y_{10}^z - \frac{2-\sqrt{2}}{4} Y_{1 - 1}^z, \\
 & Y&^{n_2}_{11}  =   \frac{2 - \sqrt{2}}{4} Y_{11}^z + \frac{i}{2}
  Y_{10}^z - \frac{ 2 + \sqrt{2}}{4} Y_{1 - 1}^z, \\
 & Y&^{n_3}_{11}  =   \frac{2 - \sqrt{2}}{4} Y_{11}^z - \frac{i}{2}
  Y_{10}^z - \frac{ 2 + \sqrt{2}}{4} Y_{1 - 1}^z, \\
 & Y&^{n_4}_{11}  =   \frac{ 2+\sqrt{2} }{4} Y_{11}^z - \frac{i}{2}
  Y_{10}^z - \frac{ 2-\sqrt{2}}{4} Y_{1 - 1}^z.
\end{eqnarray}

Because the ionizing laser is linearly polarized, the quantum number $m$ is conserved during the ionization process, which implies that the $Y^{z}_{lm}(m=0,\pm{1})$ components can be propagated independently and the final momentum distribution should be summed over these components coherently according to the above transforming coefficients. This can greatly save the computational time in comparison to directly solving the TDSE over the whole Hilbert space.

After solving the TDSE, the momentum distributions of the exited Rb atoms are averaged over the six polarization directions, i.e.,
\begin{equation}
f_{5 P} ({\bf{p}}) = \frac{1}{6}  \sum^6_{i = 1} f^{n_i}_{5 P, m =
  1} ({\bf{p}}) ,
\end{equation}
where the subscripts $5P$ and $m=1$ represent that the momentum distribution is calculated from the initial state with the orbital angular momentum $l=1$, the magnetic angular momentum $m=1$, and the superscripts indicate the polarization directions.

For the ground states $(\left|5S, m=0\right>)$, there exists a spherical symmetry, so the direction of the main axis can be chosen arbitrarily, which implies that the ionization momentum distribution from the ground state can be expressed as:
\begin{equation}
f_{5 S} ({\bf{p}}) =  f^z_{5 S, m = 0} ({\bf{p}}).
\end{equation}

Finally, the momentum distributions of the ground state and the excited state should be added together
\begin{equation}
  f ({\bf{p}}) = \frac{1}{1+\alpha} f_{5 P} ({\bf{p}}) + \frac{\alpha}{1+\alpha} f_{5 S}
  ({\bf{p}}),
\end{equation}
where $\alpha$ is the population ratio of the ground state over the excited state, which will be determined later by a best fitting between the experimental and theoretical RIMDs.

\begin{figure*}[ht]
\centering
\includegraphics[width=15.5cm,height=10cm]{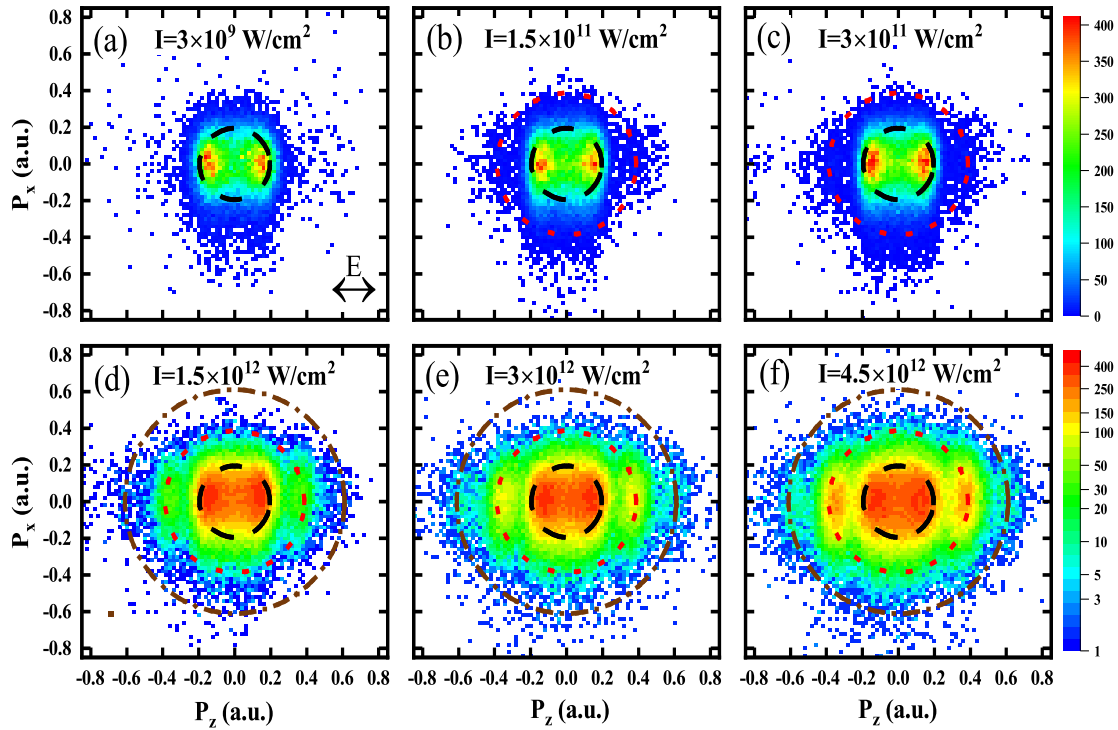}
\caption{\label{fig2}The 2D RIMDs of Rb$^+$ ions obtained in the experiment, for which $p_z$ and $p_x$ represent the momenta parallel and perpendicular to the laser polarization direction, respectively. The double arrow indicates the laser polarization direction. The laser intensities are shown on the top of each plot. The black dashed circles mark the theoretically expected recoil ion momentum from single-photon ionization of the 5$P_{3/2}$ state, while the red dotted circles and brown dashed dotted circles indicate the two-photon ionization and three-photon ionization of the 5$S_{1/2}$ state, respectively.}
\end{figure*}

\section{RESULTS AND DISCUSSION}
As described above, the Keldysh parameter $\gamma$ can be used as an indication whether tunneling picture or multi-photon ionization picture should be considered. The laser intensities in this work ranges from  $3\times10^{9}$ W/cm$^2$ to $4.5\times10^{12}$ W/cm$^2$, which correspond to  $\gamma$ values of 215.7 (169.66) to 5.57 (4.38) for the 5$S_{1/2}$ (5$P_{3/2}$) state . This allows us to consider the ionization of Rb atoms as a few photon ionization process.
It is worth noting that a certain range of the laser intensities in this work have surpassed the OBI threshold, i.e., $I_{c}=1.2\times10^{12}$ W/cm$^2$ (5$S_{1/2}$ state) and $1.77\times10^{11}$ W/cm$^2$ (5$P_{3/2}$ state), the ionization should be considered as a strong-perturbative FPI process~\cite{Nalilun,Li,Na,K}. As indicated by the vertical blue arrows in Fig.~\ref{fig1}(b), the Rb atoms in the 5$S_{1/2}$ state are mainly ionized  through two-photon ionization while those in the excited 5$P_{3/2}$ state with different magnetic quantum numbers are mainly proceed through single photon ionization. Since the photon momentum is small and can be neglected, the recoil ion momentum can be deduced from the photoelectron momentum according to the momentum conservation law, as following:
\begin{eqnarray}
&\rm{Rb}&(5P)+\hbar \omega (3.1 {\rm{eV}})\rightarrow {\rm{Rb}}^+ +e(p_r=0.2 a.u.),\\
&\rm{Rb}&(5S)+2\hbar \omega (6.2 {\rm{eV}})\rightarrow {\rm{Rb}}^+ +e(p_r=0.39 a.u.).
\end{eqnarray}
These relationships can be used as a guide to distinguish different pathways contributed by the 5$S_{1/2}$ and 5$P_{3/2}$ states to the momentum spectra.

The measured RIMDs in the polarization plane ($xz$) of the ionizing laser are presented in Fig.~\ref{fig2}. Here, to reduce the effect of diffraction associated with the laser spatial profile and its Rayleigh length, the data are extracted with constraint of the momentum along the laser propagation direction, i.e., $|p_y| < 0.2$ a.u.~\cite{cutpy,cutpy2}.
When the laser intensity is as low as $3\times10^{9}$ W/cm$^2$ [Fig.~\ref{fig2}(a)], a clear double-lobe structure can be observed, pointing to the dominance of one-photon ionization of the excited state, as indicated by the black dashed circle with the theoretically predicted momentum $p_r=0.2$ a.u. This can be understood since the photoionization cross section of the $5P_{3/2}$ state of Rb is much larger than that of the $5S_{1/2}$ state~\cite{crosssection1,crosssection2,ratio4}.
With the laser intensity increasing from $3\times10^{9}$ W/cm$^2$ to $4.5\times10^{12}$ W/cm$^2$, more structures can be observed at higher energy (momentum) part of the spectra.
For example, the two-photon ionization of Rb from the 5$S_{1/2}$ state, indicated by the red dashed circle with $p_r=0.39$ a.u. appears at $1.5\times10^{11}$ W/cm$^2$ - $4.5\times10^{12}$ W/cm$^2$.
The three-photon ATI of the 5$S_{1/2}$ state is also detected, as marked by the brown dashed dotted ring at $p_r=0.61$ a.u. in Figs.~\ref{fig2}(d)-\ref{fig2}(e). These observations demonstrate the well-known fact that the ion yields of the MPI process increases nonlinearly with the increasing laser intensity. However, the ATI peaks of the 5$P_{3/2}$ state (e.g., two photon ionization, which would be expected to show up at 0.52 a.u. if it exists) is not discernible even at the highest laser intensity.

\begin{figure}[tbh]
\centering
\includegraphics[width=8cm,height=8cm]{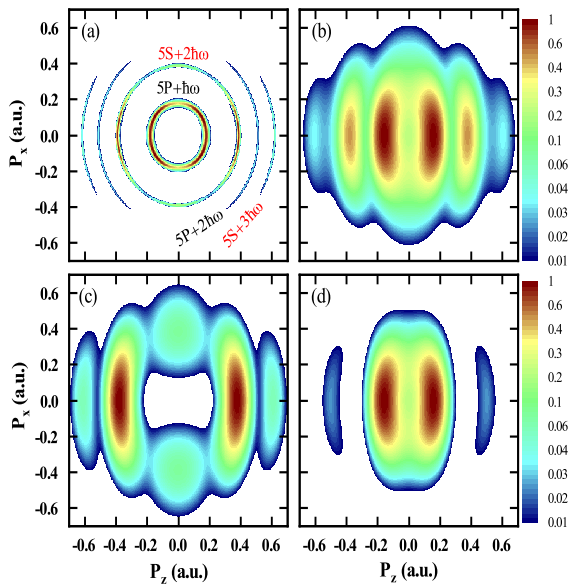}
\caption{\label{fig3}The 2D RIMDs of Rb$^+$ from TDSE simulations at $I=4.5\times10^{12}$ W/cm$^2$. The laser pulse FWHM is 35 fs, the same as that used in the experiment. (a) shows the raw data while (b) is the convolution of (a) and the resolution of the momentum measurements in the experiment, i.e., $\Delta{p}_z=0.1$ a.u. and $\Delta{p}_x=0.3$ a.u. The contributions of the ground and excited states are separated and shown in (c) and (d), respectively.}
\end{figure}

To achieve deeper insight into these experimental measurements, we then perform a series of TDSE simulations, details of which have been described in Sec. II. In general, the experimental observations are qualitatively reproduced by our quantum simulations over the whole range of laser intensities. Here, without loss of generality, we present only the data at $I=4.5\times10^{12}$ W/cm$^2$, as shown in Fig.~\ref{fig3}(a). Two main differences between theory and experiment can be readily observed: (i) the ATI ring of the 5$P_{3/2}$ state (the third ring, counting outward from the center, not observed in experiment) can be clearly identified along with three other rings observed in the experiment; and (ii) these rings are much thinner than the experimental observations. To further account for the discrepancy between experiment and theory, we then convolve the simulation results with a 2D Gaussian function to mimic the resolution of the experimental measurements. The result is shown in Fig.~\ref{fig3}(b), which is in good agreement with the experimental observation at the same laser intensity [recall Fig.~\ref{fig2}(f)]. In theory, the contributions of the ground state and the excited state to the final momentum distribution can be separated, as demonstrated in Figs.~\ref{fig3}(c) and \ref{fig3}(d). These results explain why the ATI signature of the 5$P_{3/2}$ state is barely observed in experiment, since the signal is much weaker than that of the 5$S_{1/2}$ state and, furthermore, the two-photon ATI of the 5$P_{3/2}$ state and the two-photon ionization of the 5$S_{1/2}$ state almost overlap in the momentum space due to the limited resolution of the experiment. In spite of this, the main FPI structures are retained in the convolved momentum spectra, which permits a quantitative comparison between experiments and theories as will be shown in the following.

\begin{figure}[tbh]
\centering
\includegraphics[width=8cm,height=8cm]{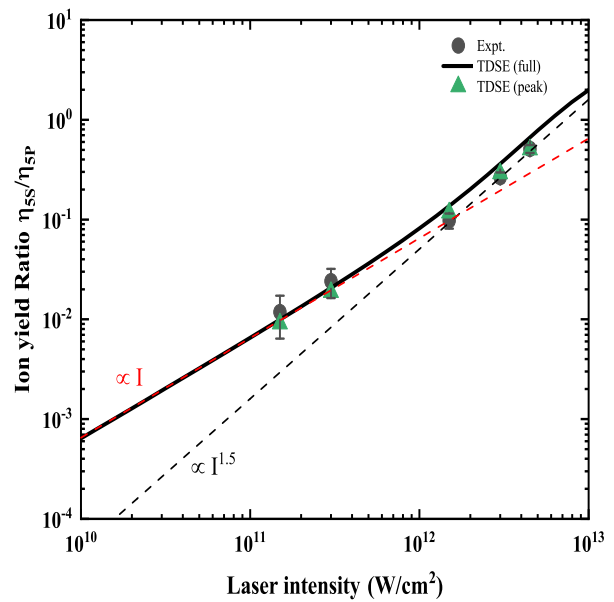}
\caption{\label{fig4}Rb$^+$ ion yield ratio of the 5$S_{1/2}$ state two-photon ionization to the 5$P_{3/2}$ state single-photon ionization. The black solid curve is the full theoretical predictions, i.e., the total ion yield after integrating over the whole momentum space, while the scatters are estimations based on the bright spots on RIMDs for the experimental data (black full circles) and the corresponding simulations (green triangles), see text for details. The best fitting between theory and experiment indicates that the population ratio of the 5$S_{1/2}$ state over the 5$P_{3/2}$ state is $\sim4:1$. The red and black dashed lines are used to guide the eyes, which demonstrates a transition of the scaling law from $I$, as predicted by a perturbative theory in the weak-field limit, to $I^{1.5}$ with the increase of $I$.}
\end{figure}

Firstly, the RIMDs provide a means of determining the population ratio $\alpha$ of Rb in the 5$S_{1/2}$ state over 5$P_{3/2}$ state.
Experimentally, $\alpha$ is determined by the cooling laser intensity and detuning~\cite{ratio1,ratio2,ratio3,ratio4}. In order to get a quantitative estimation of $\alpha$, we can take full advantage of the relative ion yields from the 5$S_{1/2}$ state and the 5$P_{3/2}$ state, represented by the ion counts (local maximum $\eta_{5S}$ and $\eta_{5P}$) at the two bright spots around the two main FPI rings shown in Figs.~\ref{fig2} and \ref{fig3}. The ratio of the ion yields $\eta_{5S}/\eta_{5P}$ is presented in Fig.~\ref{fig4} as a function of ionizing laser intensity for the experimental data (black full cicles) and the TDSE simulations (green triangles), respectively. Then $\alpha$ is treated as a parameter to be fitted with the least square method by comparing these two sets of results, and the best fitting between experiment and theory gives $\alpha=3.95$. We also note that the intensity-dependent scaling law of $\eta_{5S}/\eta_{5P}$ can serve as an indication for the onset of strong-perturbative FPI.
At lower laser intensities, according to the perturbation picture, the ion yield ratio is expected to show a slope of 1 in this double logarithmic plot, see the red dashed line in Fig.~\ref{fig4}. While for higher laser intensities, our TDSE calculation predicts a slope of about 1.5, as indicated by the black dashed line. The experimental results are in good agreement with these asymptotic predictions. The transition of the scaling law from $I$ to $I^{1.5}$ as $I$ increases is closely related to the ionization saturation of the excited state and therefore the ionization process should be considered as strong-perturbative FPI.

\begin{figure}[htbp]
\centering
\includegraphics[width=6.3cm,height=13cm]{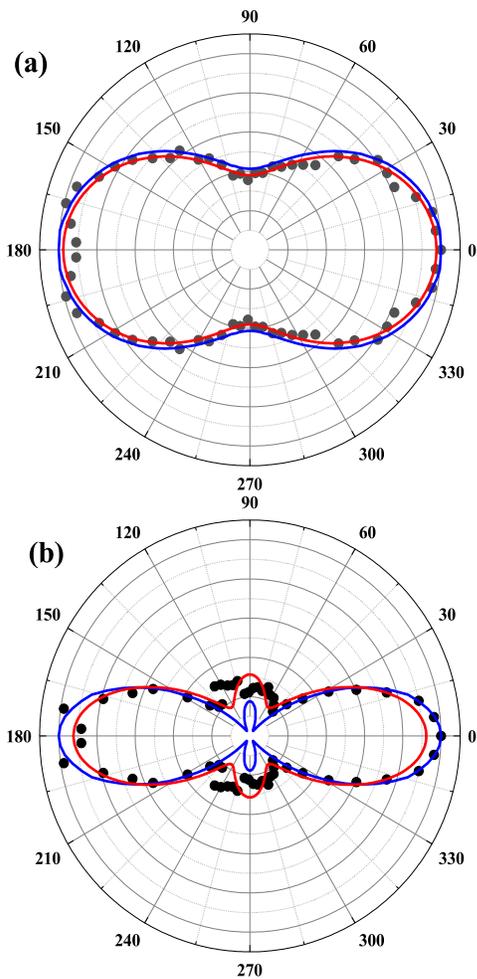}
\caption{\label{fig5}Angular distributions of the Rb$^+$ ions for single-photon ionization of the 5$P_{3/2}$ state at $I=3\times10^{9}$ W/cm$^2$ (a) and two-photon ionization of the 5$S_{1/2}$ state at $I=1.5\times10^{12}$ W/cm$^2$ (b). The scatters are the experimental measurements. The red dashed curves are fitting results of the experimental data with the Legendre polynomials, while the blue solid curves are the corresponding theoretical predictions.}
\end{figure}

\begin{figure*}[htbp]
\centering
\includegraphics[width=17cm,height=11cm]{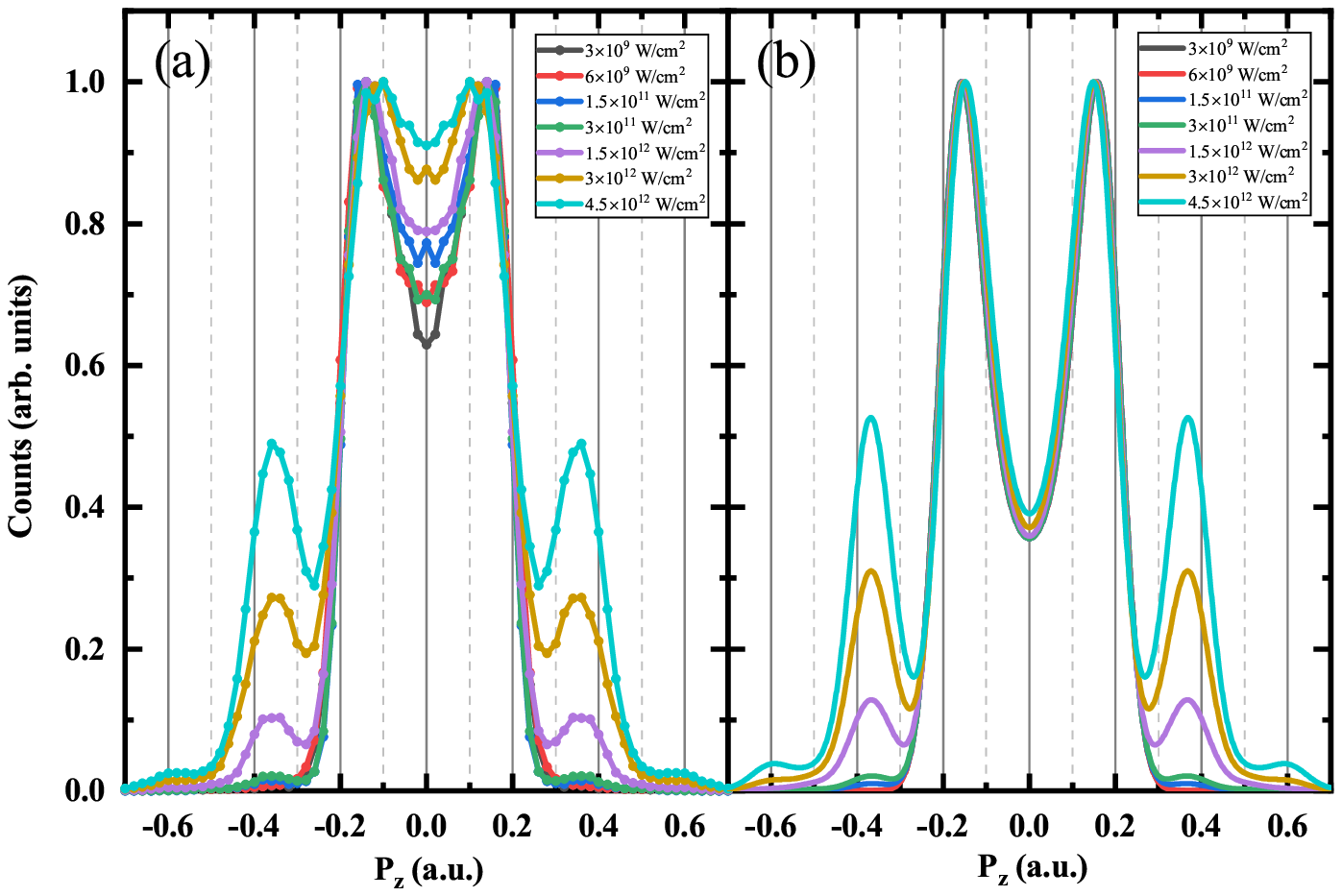}
\caption{\label{fig6}Comparison of the experimental (a) and theoretical (b) momentum distributions of the Rb$^+$ ions along the laser polarization direction. The corresponding laser intensities are indicated by the labels in the figure. Each curve is normalized independently at the peak.}
\end{figure*}

Secondly, the PAD is another sensitive probe of the strong-field atomic ionization. By integrating the data over specific momentum intervals from the 2D RIMDs, the angular distribution of the recoil ion (equivalent to the corresponding PAD as a result of momentum conservation) originated from the 5$S_{1/2}$ state and the 5$P_{3/2}$ state can be extracted separately. At $I=3\times10^{9}$ W/cm$^2$, the spectrum is dominated by one-photon ionization of the 5$P_{3/2}$ state. Therefore, the PAD has a peanut shape [see Fig.~\ref{fig5}(a)] with two main peaks at $\theta=0^{\circ}$ and $180^{\circ}$, parallel to the laser polarization direction, pointing to the dominant $d$-wave ($\left|l = 2, m= 1\right>_{n_i}$) emission according to the selection rule of dipole transitions. The result can be well fitted by the high-order Legendre polynomials:
\begin{equation}
\frac{\mathrm{d}\sigma }{\mathrm{d}\Omega }=\frac{\sigma _{0}}{4\pi}\sum_{i=0}^{n}\beta _{2i}P_{2i}(\cos \theta ),
\end{equation}
where $\sigma _{0}$ is the total photoionization cross section, $\theta$ is the angle between the photoelectron momentum vector and the polarization vector of the laser, $\beta$ is the anisotropy parameters, and $P_{2i}$ are the Legendre polynomials in variable $\cos \theta$. The fitting value of the anisotropy parameter is $\beta_2=0.69$, in good agreement with the value reported in Ref.~\cite{betaparameter} (see Fig. 5 of that paper) utilizing the configuration interaction Pauli-Fock method including core polarization potential (CIPFCP) but significantly deviate from the CIPF calculation. This pinpoints the pivotal role of the core polarization effect on the outmost electron dynamics.
With the increase of the laser intensity, the 5$S_{1/2}$ state two-photon ionization gets more and more prominent. As shown in Fig.~\ref{fig5}(b), the corresponding PAD not only shows two main peaks at $\theta=0^{\circ}$ and $180^{\circ}$ but also two side lobes in the perpendicular direction, with the fitting parameters $\beta_2=0.84$ and $\beta_4=1.13$. It is formed by a superposition of the $d$-wave ($\left|l = 2, m= 0\right>$) and $s$-wave ($\left|l = 0, m= 0\right>$). The blue curves show the TDSE simulations for comparison, which are quantitatively consistent with the experimental measurements. The present results show that the MOTRIMS can obtain high-resolution and high-quality data to provide insights into detailed structures of the final momentum spectra.

Last but not least, by integrating over $p_x$ one can obtain the momentum spectrum of Rb$^+$ along the polarization direction, as shown in Fig.~\ref{fig6}(a). Here, the peaks at $p_z$=0.15, 0.37, and 0.6 a.u. correspond to the three FPI rings in Fig.~\ref{fig2}, respectively.
For comparison, Fig.~\ref{fig6}(b) shows the results of the TDSE simulations, which are qualitatively consistent with the experimental measurements. However, it should be noted that when the laser intensity is above $3\times10^{11}$ W/cm$^2$, the experimental results show that the peak at $p_z=0.15$ a.u., corresponding to the single photon ionization of the 5$P_{3/2}$ state, moves towards momentum zero with increasing of laser intensity, which might be caused by the influence of the ponderomotive energy $U_p$, while this peak shift is not so clearly seen in theory. Another apparent difference between the experiment and theory can be observed at the dip NZM : experimentally, the depth of the dip becomes much shallower as the laser intensity increases, but it seems do not vary significantly in the theoretical simulation. We try to further explain the discrepancy by adopting another set of model potentials that include more physical effects such as the spin-orbit coupling~\cite{spinorbit} and dynamical core polarization~\cite{dynamicalcore}. However, the results are almost identical to those shown in Fig.~\ref{fig6}(b), and thus do not seem to solve the contradiction. We therefore speculate that the remained discrepancy could be induced by the dynamical electron-electron interaction that has been totally neglected in our TDSE simulations since we have adopted the single-active-electron approximation. This would be an interesting issue for further studies.

\section{SUMMARY}
With MOTRIMS technology and TDSE calculations, we investigated the single photoionization of cold rubidium atoms with the 400 nm femtosecond laser pulses, where ionization processes of the ground state 5$S_{1/2}$ and the excited-state 5$P_{3/2}$ are dominated by absorption of one-photon, two-photon, and three-photon, respectively. Experimental and theoretical RIMDs, ion yields and angular distributions are studied as a function of laser intensities at $I=3\times10^9$ W/cm$^2$ $-4.5\times10^{12}$ W/cm$^2$. With the increasing of $I$, we find that experimental NZM signals in RIMDs resulted from the 5$P_{3/2}$ state enhance dramatically and its peaked Rb$^+$ momenta dwindle obviously while that from the 5$S_{1/2}$ state is maintained. Meanwhile, the ion-yield ratio of the 5$S_{1/2}$ over the 5$P_{3/2}$ states varies from $I$ to $I^{1.5}$ as $I$ increases, where the ionization yield of 5$S_{1/2}$ state displays the intensity dependence of $I^{2}$ resulted from the two-photon absorption. These features indicate a completely perturbative two-photon ionization of 5$S_{1/2}$ state and a transition from perturbative one-photon ionization to strong-perturbative one-photon ionization of the 5$P_{3/2}$ state. The TDSE simulation reproduces the feature in the ionization curve of the 5$P_{3/2}$ state, and the simulated two-dimensional momentum distribution and angular distribution are also in good agreement with the experiment. This not only proves that the magneto-optical trap recoil ion momentum spectrometer can obtain high-resolution and high-quality data, providing insights into the detailed structure of the final momentum space, but also verifies the reliability of the simulation method using TDSE. However, there still exists some difference between the experimental results and theoretical simulations in the change of the dip NZM of one-dimensional momentum distribution with the laser intensity, indicating that there are still some details to be considered for the accurate calculation of the process. The reasons for the differences remain to be explored in the future. We hope that the work presented here will inspire further study for few-photon induced ionization processes of alkali atoms in the strong femtosecond laser field, both experimentally and theoretically.

\begin{acknowledgments}
This work is supported by the National Key R\&D Program of China (No. 2022YFA1604302) and the National Natural Science Foundation of China (NSFC)
(Grants No. 11827806 and No. 12174034). 
\end{acknowledgments}

\nocite{*}
\bibliography{apssamp}

\end{document}